# Beyond Quantifier-Free Interpolation in Extensions of Presburger Arithmetic (Extended Technical Report)*

Angelo Brillout[1], Daniel Kroening[2], Philipp Rümmer[2], and Thomas Wahl[2]

[1] ETH Zurich, Switzerland
[2] Oxford University Computing Laboratory, United Kingdom

**Abstract.** Craig interpolation has emerged as an effective means of generating candidate program invariants. We present interpolation procedures for the theories of Presburger arithmetic combined with (i) uninterpreted predicates (QPA+UP), (ii) uninterpreted functions (QPA+UF) and (iii) extensional arrays (QPA+AR). We prove that none of these combinations can be effectively interpolated without the use of quantifiers, even if the input formulae are quantifier-free. We go on to identify fragments of QPA+UP and QPA+UF with restricted forms of *guarded quantification* that are closed under interpolation. Formulae in these fragments can easily be mapped to quantifier-free expressions with integer division. For QPA+AR, we formulate a sound interpolation procedure that potentially produces interpolants with unrestricted quantifiers.

## 1 Introduction

Given two first-order logic formulae $A$ and $C$ such that $A$ implies $C$, written $A \Rightarrow C$, *Craig interpolation* determines a formula $I$ such that the implications $A \Rightarrow I$ and $I \Rightarrow C$ hold, and $I$ contains only non-logical symbols occurring in both $A$ and $C$ [2]. Interpolation has emerged as a practical approximation method in computing and has found many uses in formal verification, ranging from efficient image computations in SAT-based model checking, to computing candidate invariants in automated program analysis.

In software verification, interpolation is applied to formulae encoding the transition relation of a model underlying the program. In order to support a wide variety of programming language constructs, much effort has been invested in the design of algorithms that compute interpolants for formulae of various first-order theories. For example, interpolating integer arithmetic solvers have been reported for fragments such as difference-bound logic, linear equalities, and constant-divisibility predicates.

The goal of this paper is an interpolation procedure that is instrumental in analysing programs manipulating integer variables. We therefore consider the first-order theory of *quantified Presburger arithmetic* (quantified linear integer arithmetic), denoted QPA. Combined with *uninterpreted predicates* (UP) and *uninterpreted functions* (UF), this allows us to encode the theory of *extensional arrays* (AR), using uninterpreted function symbols for read and write operations. Our interpolation procedure extracts an interpolant directly from a proof of $A \Rightarrow C$. Starting from a sound and complete proof system based on a sequent calculus, the proof rules are extended by labelled formulae and annotations that reduce, at the root of a closed proof, to interpolants. In earlier work, we presented a similar procedure for quantifier-free Presburger arithmetic [3].

* This research is supported by the EPSRC project EP/G026254/1, by the EU FP7 STREP MOGENTES, and by the EU ARTEMIS CESAR project. A peer-refereed version of this report is available at `www.springerlink.com` [1].

In program verification, an interpolating theorem prover often interacts tightly with various decision procedures. It is therefore advantageous for the interpolants computed by the prover to be expressible in simple logic fragments. Unfortunately, interpolation procedures for expressive first-order fragments, such as integer arithmetic with uninterpreted predicates, often generate interpolants with *quantifiers*, which makes subsequent calls to decision procedures involving these interpolants expensive. This is not by accident. In fact, in this paper we first show that interpolation of QPA+UP in general requires the use of quantifiers, *even if the input formulae are themselves free of quantifiers.*

In order to solve this problem, we study fragments of QPA+UP that are *closed under interpolation*: fragments such that interpolants for input formulae can again be expressed in the theory. By the result above, such fragments must allow at least a limited form of quantification. Our second contribution is to show that the theory PAID+UP of Presburger arithmetic with uninterpreted predicates and a restricted form of *guarded quantifiers* indeed has the closure property. A similar fragment, PAID+UF, can be identified for the combination of Presburger arithmetic with uninterpreted functions. Moreover, by allowing *integer divisibility* (ID) predicates, the guarded quantifiers can be rewritten into quantifier-free form, facilitating further processing of the interpolants.

In summary, we present in this paper an interpolating calculus for the first-order theory of Presburger arithmetic and uninterpreted predicates, QPA+UP. We show that, for some quantifier-free input formulae, quantifiers in interpolants cannot be avoided, and suggest a restriction of QPA+UP that is closed under interpolation, yet permits quantifier-free interpolants conveniently expressible in standard logics. We extend these results to Presburger theories with uninterpreted functions and, specifically, to quantified array theory, resulting in the first sound interpolating decision procedure for Presburger arithmetic and arrays.

## 2 Background

### 2.1 Presburger Arithmetic with Predicates and Functions

*Presburger arithmetic.* We assume familiarity with classical first-order logic (e.g., [4]). Let $x$ range over an infinite set $X$ of variables, $c$ over an infinite set $C$ of constants, $p$ over a set $P$ of uninterpreted predicates with fixed arity, $f$ over a set $F$ of uninterpreted functions with fixed arity, and $\alpha$ over the set $\mathbb{Z}$ of integers. (Note the distinction between constant *symbols*, such as $c$, and integer *literals*, such as 42.) The syntax of terms and formulae considered in this paper is defined by the following grammar:

$$\begin{aligned}\phi \ &::= \ t \doteq 0 \ \big| \ t \leq 0 \ \big| \ \alpha \mid t \ \big| \ p(t, \ldots, t) \ \big| \ \phi \wedge \phi \ \big| \ \phi \vee \phi \ \big| \ \neg\phi \ \big| \ \forall x.\phi \ \big| \ \exists x.\phi \\ t \ &::= \ \alpha \ \big| \ c \ \big| \ x \ \big| \ \alpha t + \cdots + \alpha t \ \big| \ f(t, \ldots, t)\end{aligned}$$

The symbol $t$ denotes terms of linear arithmetic. Divisibility atoms $\alpha \mid t$ are equivalent to formulae $\exists s.\ \alpha s - t \doteq 0$, but are required for quantifier-free interpolation. Simultaneous substitution of a vector of terms $\bar{t} = (t_1, \ldots, t_n)$ for variables $\bar{x} = (x_1, \ldots, x_n)$ in $\phi$ is denoted by $[\bar{x}/\bar{t}]\phi$; we assume that variable capture is avoided by renaming bound variables as necessary. For simplicity, we sometimes write $s \doteq t$ as a shorthand of $s - t \doteq 0$, and $\forall c.\phi$ as a shorthand of $\forall x.[c/x]\phi$ if $c$ is a constant. The abbreviation *true* (*false*) stands for the equality $0 \doteq 0$ ($1 \doteq 0$), and the formula $\phi \rightarrow \psi$ abbreviates $\neg\phi \vee \psi$. Semantic notions such as structures, models, satisfiability, and validity are defined as is common over the universe $\mathbb{Z}$ of integers (e.g., [4]).

Full *quantified Presburger arithmetic* (QPA) consists of the formulae that do not contain uninterpreted predicates or functions; *(quantifier-free) Presburger arithmetic* (PA) is the quantifier-free fragment of QPA. The logic QPA+UP (QPA+UF) extends QPA to formulae with uninterpreted predicates (functions), according to the above grammar.

### 2.2 An Interpolating Sequent Calculus

*Interpolating sequents.* To extract interpolants from unsatisfiability proofs of $A \wedge B$, formulae are labelled either with the letter $L$ ("left") to indicate that they are derived from $A$ or with $R$ ("right") for formulae derived from $B$ (as in [3]). More formally, if $\phi$ is a formula without free variables, then $\lfloor\phi\rfloor_L$ and $\lfloor\phi\rfloor_R$ are $L/R$-labelled formulae. If $\Gamma$, $\Delta$ are finite sets of labelled formulae and $I$ is an unlabelled formula without free variables, then $\Gamma \vdash \Delta \blacktriangleright I$ is an *interpolating sequent*. Similarly, if $\Gamma$, $\Delta$ are sets of unlabelled formulae without free variables, then $\Gamma \vdash \Delta$ is an (ordinary) *sequent*. An ordinary sequent is *valid* if the formula $\bigwedge\Gamma \rightarrow \bigvee\Delta$ is valid.

The semantics of interpolating sequents is defined using the projections $\Gamma_L =_{\text{def}} \{\phi \mid \lfloor\phi\rfloor_L \in \Gamma\}$ and $\Gamma_R =_{\text{def}} \{\phi \mid \lfloor\phi\rfloor_R \in \Gamma\}$, which extract the $L/R$-parts of a set $\Gamma$ of labelled formulae. A sequent $\Gamma \vdash \Delta \blacktriangleright I$ is *valid* if (i) the sequent $\Gamma_L \vdash \Delta_L, I$ is valid, (ii) the sequent $\Gamma_R, I \vdash \Delta_R$ is valid, and (iii) the constants and uninterpreted predicate/functions in $I$ occur in both $\Gamma_L \cup \Delta_L$ and $\Gamma_R \cup \Delta_R$. As special cases, $\lfloor A\rfloor_L \vdash \lfloor C\rfloor_R \blacktriangleright I$ reduces to $I$ being an interpolant of the implication $A \Rightarrow C$, while $\lfloor A\rfloor_L, \lfloor B\rfloor_R \vdash \emptyset \blacktriangleright I$ captures the concept of interpolants for unsatisfiable conjunctions $A \wedge B$ common in formal verification.

*Interpolating sequent calculi.* An *interpolating rule* is a binary relation between a finite set of interpolating sequents, called the premises, and a sequent called

$$\frac{\Gamma, \lfloor \phi \rfloor_L \vdash \Delta \blacktriangleright I \qquad \Gamma, \lfloor \psi \rfloor_L \vdash \Delta \blacktriangleright J}{\Gamma, \lfloor \phi \vee \psi \rfloor_L \vdash \Delta \blacktriangleright I \vee J} \text{ OR-LEFT-L} \qquad \frac{\Gamma, \lfloor \phi \rfloor_R \vdash \Delta \blacktriangleright I \qquad \Gamma, \lfloor \psi \rfloor_R \vdash \Delta \blacktriangleright J}{\Gamma, \lfloor \phi \vee \psi \rfloor_R \vdash \Delta \blacktriangleright I \wedge J} \text{ OR-LEFT-R}$$

$$\frac{\Gamma, \lfloor \phi \rfloor_D, \lfloor \psi \rfloor_D \vdash \Delta \blacktriangleright I}{\Gamma, \lfloor \phi \wedge \psi \rfloor_D \vdash \Delta \blacktriangleright I} \text{ AND-LEFT} \qquad \frac{\Gamma \vdash \lfloor \phi \rfloor_D, \Delta \blacktriangleright I}{\Gamma, \lfloor \neg\phi \rfloor_D \vdash \Delta \blacktriangleright I} \text{ NOT-LEFT}$$

$$\frac{*}{\Gamma, \lfloor \phi \rfloor_L \vdash \lfloor \phi \rfloor_L, \Delta \blacktriangleright \mathit{false}} \text{ CLOSE-LL} \qquad \frac{*}{\Gamma, \lfloor \phi \rfloor_R \vdash \lfloor \phi \rfloor_R, \Delta \blacktriangleright \mathit{true}} \text{ CLOSE-RR}$$

$$\frac{*}{\Gamma, \lfloor \phi \rfloor_L \vdash \lfloor \phi \rfloor_R, \Delta \blacktriangleright \phi} \text{ CLOSE-LR} \qquad \frac{*}{\Gamma, \lfloor \phi \rfloor_R \vdash \lfloor \phi \rfloor_L, \Delta \blacktriangleright \neg\phi} \text{ CLOSE-RL}$$

$$\frac{\Gamma, \lfloor [x/t]\phi \rfloor_L, \lfloor \forall x.\phi \rfloor_L \vdash \Delta \blacktriangleright I}{\Gamma, \lfloor \forall x.\phi \rfloor_L \vdash \Delta \blacktriangleright \forall_{Rt}\, I} \text{ ALL-LEFT-L} \qquad \frac{\Gamma, \lfloor [x/t]\phi \rfloor_R, \lfloor \forall x.\phi \rfloor_R \vdash \Delta \blacktriangleright I}{\Gamma, \lfloor \forall x.\phi \rfloor_R \vdash \Delta \blacktriangleright \exists_{Lt}\, I} \text{ ALL-LEFT-R}$$

$$\frac{\Gamma, \lfloor [x/c]\phi \rfloor_D \vdash \Delta \blacktriangleright I}{\Gamma, \lfloor \exists x.\phi \rfloor_D \vdash \Delta \blacktriangleright I} \text{ EX-LEFT} \qquad \frac{\Gamma \vdash \lfloor [x/c]\phi \rfloor_D, \Delta \blacktriangleright I}{\Gamma \vdash \lfloor \forall x.\phi \rfloor_D, \Delta \blacktriangleright I} \text{ ALL-RIGHT}$$

**Fig. 1.** The upper box presents a selection of interpolating rules for propositional logic, while the lower box shows the interpolating rules to handle quantifiers. Parameter $D$ stands for either $L$ or $R$. The quantifier $\forall_{Rt}$ denotes universal quantification over all constants occurring in $t$ but not in $\Gamma_L \cup \Delta_L$; likewise, $\exists_{Lt}$ denotes existential quantification over all constants occurring in $t$ but not in $\Gamma_R \cup \Delta_R$. In the rules EX-LEFT and ALL-RIGHT, $c$ is a constant that does not occur in the conclusion.

the conclusion:

$$\frac{\Gamma_1 \vdash \Delta_1 \blacktriangleright I_1 \quad \cdots \quad \Gamma_n \vdash \Delta_n \blacktriangleright I_n}{\Gamma \vdash \Delta \blacktriangleright I}$$

An interpolating rule is *sound* if, for all instances whose premises $\Gamma_1 \vdash \Delta_1 \blacktriangleright I_1$, $\ldots$, $\Gamma_n \vdash \Delta_n \blacktriangleright I_n$ are valid, the conclusion $\Gamma \vdash \Delta \blacktriangleright I$ is valid, too. Fig. 1 presents a selection of interpolating rules (used throughout the paper) for predicate logic. An exhaustive list of rules is given in Appendix A (Fig. 4) and in [3].

Interpolating proofs are trees growing upwards, in which each node is labelled with an interpolating sequent, and each non-leaf node is related to the node(s) directly above it through an instance of a calculus rule. A proof is *closed* if it is finite and all leaves are justified by an instance of a rule without premises.

To construct a proof for an interpolation problem, we build a proof tree starting from the root $\Gamma \vdash \Delta \blacktriangleright I$ with unknown interpolant $I$, i.e., $I$ acts as a place holder. For example, to solve an interpolation problem $A \wedge B$, we start with the sequent $\lfloor A \rfloor_L, \lfloor B \rfloor_R \vdash \emptyset \blacktriangleright I$. Rules are then applied successively to decompose and simplify the sequent. Once all branches are closed, i.e., a proof is found, an interpolant can be extracted from the proof. Starting from the leaves, intermediate interpolants are computed and propagated back to the root leading to an interpolant $I$. An example of this procedure is given in the next section.

## 3 Interpolation for Uninterpreted Predicates

### 3.1 Presburger Arithmetic and Uninterpreted Predicates

We begin by studying the interpolation problem for Presburger arithmetic extended with uninterpreted predicates (QPA+UP), which forms a simple yet expressive base logic in which functions and arrays can be elegantly encoded. The case of predicates is instructive, since essentially the same phenomena occur under interpolation as with uninterpreted functions.

*Example 1.* We illustrate the construction of an interpolating proof by deriving an interpolant for $A \Rightarrow C$, with $A = (\neg p(c) \vee p(d)) \wedge p(c)$ and $C = p(d)$. A complete interpolating proof of this implication looks as follows:

$$\dfrac{\dfrac{\dfrac{\dfrac{*}{\lfloor p(c)\rfloor_L \;\vdash\; \lfloor p(d)\rfloor_R, \lfloor p(c)\rfloor_L \;\blacktriangleright \mathit{false}}\,\textsc{close-ll}}{\lfloor \neg p(c)\rfloor_L, \lfloor p(c)\rfloor_L \;\vdash\; \lfloor p(d)\rfloor_R \;\blacktriangleright \mathit{false}}\,\textsc{not-left} \qquad \dfrac{*}{\lfloor p(d)\rfloor_L, \lfloor p(c)\rfloor_L \;\vdash\; \lfloor p(d)\rfloor_R \;\blacktriangleright p(d)}\,\textsc{close-lr}}{\lfloor \neg p(c) \vee p(d)\rfloor_L, \lfloor p(c)\rfloor_L \;\vdash\; \lfloor p(d)\rfloor_R \;\blacktriangleright \mathit{false} \vee p(d)}\,\textsc{or-left-l}}{\lfloor (\neg p(c) \vee p(d)) \wedge p(c)\rfloor_L \;\vdash\; \lfloor p(d)\rfloor_R \;\blacktriangleright \mathit{false} \vee p(d)}\,\textsc{and-left}$$

The shaded regions indicate the parts of the formula being matched against the rules in Fig. 1. The sequent $\lfloor(p(c) \vee p(d)) \wedge p(c)\rfloor_L \;\vdash\; \lfloor p(d)\rfloor_R \;\blacktriangleright I$ is the root of the proof, where $I = \mathit{false} \vee p(d)$ has been filled in once the proof was closed. The AND-LEFT rule propagates the $L$-label to the subformulae of the antecedent of the first sequent. By applying OR-LEFT-L to the disjunction $p(c) \vee p(d)$, the proof splits into two branches. The right branch can immediately be closed using CLOSE-LR. The left branch requires an application of NOT-LEFT before it can be closed with CLOSE-LL. We compute an interpolant by propagating (intermediate) interpolants from the leaves back to root of the proof. As specified by CLOSE-LR, the interpolant of the right branch is $p(d)$. On the left branch, the CLOSE-LL rule yields the interpolant *false*, which is carried through by NOT-LEFT. The rule OR-LEFT-L takes the interpolants of its two subproofs and generates $\mathit{false} \vee p(d)$. This is the final interpolant, since the last rule AND-LEFT propagates interpolants without applying modifications. □

In this example, the arguments of occurrences of uninterpreted predicates literally matched up, which need not be the case. The rules presented so far are insufficient to prove more complex theorems, such as $p(c) \wedge c \doteq d \rightarrow p(d)$, in which arithmetic and predicate calculus interact. To fully integrate uninterpreted predicates, we use an explicit *predicate consistency* axiom

$$PC_p = \quad \forall \bar{x}, \bar{y}.\ \big((p(\bar{x}) \;\wedge\; \bar{x} - \bar{y} \doteq 0) \;\rightarrow\; p(\bar{y})\big) \tag{1}$$

which can be viewed as an $L$- or $R$-labelled formula that is implicitly present in every sequent. The label $L/R$ is chosen depending on whether $p$ occurs in $\Gamma_L \cup \Delta_L$, in $\Gamma_R \cup \Delta_R$, or in both.

To make use of (1) in a proof, we need additional proof rules to instantiate quantifiers, which are given in the bottom part of Fig. 1. Formula (1) can be instantiated with techniques similar to the e-matching in SMT solvers [5]: it suffices to generate a ground instance of (1) by applying ALL-LEFT-L/R whenever literals $p(\bar{s})$ and $p(\bar{t})$ occur in the antecedent and succedent [19]:

$$\frac{\Gamma, \lfloor p(\bar{s}) \rfloor_D, \lfloor (p(\bar{s}) \ \wedge \ \bar{s} - \bar{t} \doteq 0) \ \rightarrow \ p(\bar{t}) \rfloor_L \ \vdash \ \lfloor p(\bar{t}) \rfloor_E, \Delta \ \blacktriangleright I}{\Gamma, \lfloor p(\bar{s}) \rfloor_D \ \vdash \ \lfloor p(\bar{t}) \rfloor_E, \Delta \ \blacktriangleright \forall_{R\bar{s}\bar{t}}\, I} \ \textsc{all-left-l}^+$$

where $D, E \in \{L, R\}$ are arbitrary labels, and $\forall_{R\bar{s}\bar{t}}$ denotes universal quantification over all constants occurring in the terms $\bar{s}, \bar{t}$ but not in the set of left formulae $\big(\Gamma, \lfloor p(\bar{s}) \rfloor_D\big)_L \cup \big(\Delta, \lfloor p(\bar{t}) \rfloor_E\big)_L$ (like in Fig. 1). Similarly, instances of (1) labelled with $R$ can be generated using ALL-LEFT-R. To improve efficiency, refinements can be formulated that drastically reduce the number of generated instances [7].

*Correctness.* The calculus consisting of the rules in Fig. 1, the arithmetic rules of [3], and axiom (1) generates correct interpolants. That is, whenever a sequent $\lfloor A \rfloor_L \ \vdash \ \lfloor C \rfloor_R \ \blacktriangleright I$ is derived, the implications $A \Rightarrow I$ and $I \Rightarrow C$ are valid, and the constants and predicates in $I$ occur in both $A$ and $C$. More precisely:

**Lemma 2 (Soundness).** *If an interpolating QPA+UP sequent $\Gamma \ \vdash \ \Delta \ \blacktriangleright I$ is provable in the calculus, then it is valid.*

In particular, the sequent $\Gamma_L, \Gamma_R \ \vdash \ \Delta_L, \Delta_R$ is valid in this case. As shown in [3], Lem. 2 holds for the calculus consisting of the arithmetic and propositional rules. It is easy to see that the additional rules presented in this paper are sound, too.

Concerning completeness, we observe that the logic of quantified Presburger arithmetic with predicates is $\Pi^1_1$-complete, which means that no complete calculi exist [8]. On the next pages, we therefore discuss how to restrict the quantification allowed in formulae to achieve completeness, while retaining the ability to extract interpolants from proofs.

### 3.2 Quantifiers in QPA+UP Interpolants

We first consider the quantifier-free fragment PA+UP. With the help of results in [19, 3], it is easy to see that our calculus is sound and complete for PA+UP, and can in fact be turned into a decision procedure. There is a caveat, however: although formulae in PA+UP are quantifier-free, generated interpolants may still contain quantifiers and thus lie outside of PA+UP. The source of quantifiers are the rules ALL-LEFT-L/R in Fig. 1, which can be used to instantiate $L/R$-labelled quantified formulae with terms containing alien symbols. Such symbols have to be eliminated from resulting interpolants through quantifiers. The following example illustrates this situation.

*Example 3.* Fig. 2 shows the derivation of an interpolant for the unsatisfiable conjunction $\big(2c - y \doteq 0 \wedge p(c)\big) \wedge \big(2d - y \doteq 0 \wedge \neg p(d)\big)$. After propositional reductions, we instantiate $PC_p$ with the predicate arguments $c$ and $d$, due to the occurrences of the literals $p(c)$ and $p(d)$ in the sequent. The proof can then be

$$\frac{*}{\ldots, \lfloor 2c - y \doteq 0 \rfloor_L, \lfloor 2d - y \doteq 0 \rfloor_R \ \vdash\ \lfloor c - d \doteq 0 \rfloor_L, \ldots \ \blacktriangleright y - 2d \not\doteq 0}$$

$$\mathcal{D}$$

$$\cfrac{\cfrac{\cfrac{\cfrac{\cfrac{\cfrac{*}{\ldots, \lfloor p(c) \rfloor_L \ \vdash\ \lfloor p(c) \rfloor_L \ \blacktriangleright \mathit{false}} \qquad \mathcal{D} \qquad \cfrac{*}{\ldots, \lfloor p(d) \rfloor_L \ \vdash\ \lfloor p(d) \rfloor_R \ \blacktriangleright p(d)}}{\ldots, \lfloor (p(c) \land c - d \doteq 0) \to p(d) \rfloor_L \ \vdash\ \ldots \ \blacktriangleright y - 2d \not\doteq 0 \lor p(d)} \text{ OR-LEFT-L}^+}{\lfloor PC_p \rfloor_L, \lfloor PC_p \rfloor_R, \lfloor p(c) \rfloor_L, \lfloor 2c - y \doteq 0 \rfloor_L, \lfloor 2d - y \doteq 0 \rfloor_R \ \vdash\ \lfloor p(d) \rfloor_R \ \blacktriangleright I} \text{ ALL-LEFT-L}}{\lfloor PC_p \rfloor_L, \lfloor PC_p \rfloor_R, \lfloor p(c) \rfloor_L, \lfloor 2c - y \doteq 0 \rfloor_L, \lfloor 2d - y \doteq 0 \rfloor_R, \lfloor \neg p(d) \rfloor_R \ \vdash\ \ \blacktriangleright I} \text{ NOT-LEFT}}{\lfloor PC_p \rfloor_L, \lfloor PC_p \rfloor_R, \lfloor p(c) \rfloor_L, \lfloor 2c - y \doteq 0 \rfloor_L, \lfloor 2d - y \doteq 0 \land \neg p(d) \rfloor_R \ \vdash\ \ \blacktriangleright I} \text{ AND-LEFT}}{\lfloor PC_p \rfloor_L, \lfloor PC_p \rfloor_R, \lfloor 2c - y \doteq 0 \land p(c) \rfloor_L, \lfloor 2d - y \doteq 0 \land \neg p(d) \rfloor_R \ \vdash\ \ \blacktriangleright I} \text{ AND-LEFT}$$

**Fig. 2.** Example proof involving uninterpreted predicates.

closed using propositional rules, complementary literals, and arithmetic reasoning [3]. The final interpolant is the formula $I = \forall x.\ (y - 2x \not\doteq 0 \lor p(x))$, in which a quantifier has been introduced via ALL-LEFT-L to eliminate the constant $d$. □

In fact, as we formally prove in Appendix B, quantifier-free interpolants for the inconsistent PA+UP formulae $2c - y \doteq 0 \land p(c)$ and $2d - y \doteq 0 \land \neg p(d)$ do not exist. Abstracting from this example, we obtain:

**Theorem 4.** *PA+UP is not closed under interpolation.*

Intuitively, Theorem 4 holds because the logic PA does not provide an integer division operator. Divisibility predicates $\alpha \mid t$ are insufficient in the presence of uninterpreted predicates, because they cannot be used within terms: no quantifier-free formula can express the statement $\forall x.\ (y - 2x \not\doteq 0 \lor p(x))$, which is equivalent to $2 \mid y \to p(\frac{y}{2})$.

Adding integer division is sufficient to close PA+UP under interpolation. More formally, we define the logic PAID ("PA with Integer Divisibility"), extending PA by *guarded* quantified expressions

$$\forall x.\ (\alpha x + t \not\doteq 0 \lor \phi), \qquad \exists x.\ (\alpha x + t \doteq 0 \land \phi) \tag{2}$$

where $x \in X$ ranges over variables, $\alpha \in \mathbb{N} \setminus \{0\}$ over non-zero integers, $t$ over terms not containing $x$, and $\phi$ over PAID formulae (possibly containing $x$ as a free variable). The logic PAID+UP is obtained by adding uninterpreted predicates to PAID. Note that the interpolant $I$ computed in Example 3 is in PAID+UP.

It is easy to extend our interpolating calculus to a sound and complete calculus for PAID+UP; the only necessary additional rules are

$$\frac{\Gamma, \lfloor (\alpha \nmid t) \lor \exists x.\ (\alpha x + t \doteq 0 \land \phi) \rfloor_D \ \vdash\ \Delta \ \blacktriangleright I}{\Gamma, \lfloor \forall x.\ (\alpha x + t \not\doteq 0 \lor \phi) \rfloor_D \ \vdash\ \Delta \ \blacktriangleright I} \text{ ALL-LEFT-GRD}$$

$$\frac{\Gamma \ \vdash\ \lfloor (\alpha \mid t) \land \forall x.\ (\alpha x + t \not\doteq 0 \lor \phi) \rfloor_D, \Delta \ \blacktriangleright I}{\Gamma \ \vdash\ \lfloor \exists x.\ (\alpha x + t \doteq 0 \land \phi) \rfloor_D, \Delta \ \blacktriangleright I} \text{ EX-RIGHT-GRD}$$

with the side conditions that $\alpha \neq 0$, and that $x$ does not occur in $t$.

**Theorem 5 (Completeness).** *Suppose $\Gamma, \Delta$ are sets of labelled PAID+UP formulae. If the sequent $\Gamma_L, \Gamma_R \vdash \Delta_L, \Delta_R$ is valid, then there is a formula $I$ such that (i) the sequent $\Gamma \vdash \Delta \blacktriangleright I$ is provable in the calculus of Sect. 3.1, enriched with the rules* ALL-LEFT-GRD *and* EX-RIGHT-GRD*, and (ii) $I$ is a PAID+UP formula up to normalisation of guards to obtain expressions of the form* (2)*.*

Guard normalisation is necessary in general, because interpolants generated by proofs can take the shape $\forall \bar{x}.\ (t_1 \not\doteq 0 \vee \cdots \vee t_k \not\doteq 0 \vee \phi)$, grouping together multiple quantifiers and guards. We show in Appendix C.1 that such formulae can effectively be transformed to the form (2). To prove the theorem, we first argue that sequent proofs of a certain restricted form are guaranteed to result in PAID+UP interpolants, up to normalisation of guards:

**Lemma 6.** *Suppose that every instantiation of the axiom* (1) *in a proof $\mathcal{P}$ of the PAID+UP sequent $\Gamma \vdash \Delta \blacktriangleright I$ has the form*

$$
\dfrac{\dfrac{\dfrac{*}{\ldots, \lfloor p(\bar{s}) \rfloor_D \vdash \lfloor p(\bar{s}) \rfloor_F, \ldots \blacktriangleright J_1} \quad \dfrac{\dfrac{\vdots}{\ldots, \lfloor p(\bar{s}) \rfloor_D \vdash \lfloor \bar{s} - \bar{t} \doteq 0 \rfloor_F, \lfloor p(\bar{t}) \rfloor_E, \ldots \blacktriangleright J_2}}{\mathcal{Q}} \quad \dfrac{*}{\ldots, \lfloor p(\bar{t}) \rfloor_F \vdash \lfloor p(\bar{t}) \rfloor_E, \ldots \blacktriangleright J_3}}{\ldots, \lfloor (p(\bar{s}) \wedge \bar{s} - \bar{t} \doteq 0) \to p(\bar{t}) \rfloor_F \vdash \ldots \blacktriangleright J_4}\ \text{OR-LEFT}^+}{\ldots, \lfloor p(\bar{s}) \rfloor_D \vdash \lfloor p(\bar{t}) \rfloor_E, \ldots \blacktriangleright J_5}\ \text{ALL-LEFT}^+
$$

*where (i) $D, E \in \{L, R\}$ and $F \in \{D, E\}$ are arbitrary labels, (ii) the proof $\mathcal{Q}$ only uses the rules* RED-RIGHT*,* MUL-RIGHT*,* IPI-RIGHT*,* AND-RIGHT-L*, and* CLOSE-EQ-RIGHT *applied to an equality derived from $\bar{s} - \bar{t} \doteq 0$ (see [3] or Appendix A for definitions of the rules), and (iii)* ALL-LEFT *and* EX-RIGHT *are not applied in any other places in $\mathcal{P}$. Then $I$ is a PAID+UP formula up to normalisation of guards.*

A proof of this lemma is contained in Appendix C.1. Intuitively, the conditions in the lemma enable the application of (1) to atoms $p(\bar{s})$ and $p(\bar{t})$ only if the equations present in a sequent entail that the arguments $\bar{s}$ and $\bar{t}$ match up. There are various ways of relaxing this restriction: most importantly, the applications of axiom (1) only has to be constrained when unifying literals $\lfloor p(\bar{s}) \rfloor_D$ and $\lfloor p(\bar{t}) \rfloor_E$ with *distinct* labels $D \neq E$. Applications of the axiom to literals with the same label are uncritical, because they never introduce quantifiers in interpolants. In fact, practical experience with our theorem prover PRINCESS shows that generated interpolants are often naturally in the PAID+UP fragment, even when not imposing any restrictions on the proof generation process.

The second ingredient in proving the completeness theorem Thm. 5 is to show that the calculus with the restrictions imposed in Lem. 6 is still complete. We describe a proof procedure abiding by these restrictions in Appendix C.2. As a corollary of the completeness, we obtain:

**Corollary 7.** *PAID+UP is closed under interpolation.*

Despite this closure property, some proofs may result in interpolants outside PAID+UP, by applying "wrong" rules in the sub-proof $\mathcal{Q}$ of Lem. 6:

*Example 8.* Starting from PAID+UP input formulae, the following proof generates the interpolant $\forall c.\, p(c)$, which is not equivalent to any PAID+UP formula:

$$\dfrac{\dfrac{\dfrac{*}{\lfloor p(0) \rfloor_L \ \vdash\ \lfloor p(0) \rfloor_L \ \blacktriangleright \mathit{false}} \quad \dfrac{*}{\lfloor q \rfloor_L \ \vdash\ \lfloor c \doteq 0 \rfloor_L, \lfloor q \rfloor_L \ \blacktriangleright \mathit{false}} \quad \dfrac{*}{\lfloor p(c) \rfloor_L \ \vdash\ \lfloor p(c) \rfloor_R \ \blacktriangleright p(c)}}{\ldots, \lfloor p(0) \rfloor_L, \lfloor q \rfloor_L, \lfloor (p(0) \wedge c \doteq 0) \to p(c) \rfloor_L \ \vdash\ \lfloor c \rfloor_R, \lfloor q \rfloor_L \ \blacktriangleright p(c)}}{\lfloor PC_p \rfloor_L, \lfloor PC_p \rfloor_R, \lfloor p(0) \rfloor_L, \lfloor q \rfloor_L \ \vdash\ \lfloor p(c) \rfloor_R, \lfloor q \rfloor_L \ \blacktriangleright \forall c.\, p(c)} \ \textsc{all-left-l}$$

The first step in the proof is to instantiate axiom (1), in an attempt to unify the formula $\lfloor p(0) \rfloor_L$ and $\lfloor p(c) \rfloor_R$; this instantiation later introduces the unguarded quantifier $\forall c$ in the interpolant. The proof violates the conditions in Lem. 6, because the middle sub-proof is closed using the atoms $\lfloor q \rfloor_L$ instead of the equation $\lfloor c \doteq 0 \rfloor_L$. A correct PAID+UP interpolant for this example is *false*. □

*PAID and integer division.* Despite the presence of guarded quantifiers, PAID is close to simple quantifier-free assertion languages found in programming languages like Java or C, making PAID expressions convenient to pass on to decision procedures. Specifically, the following equivalences hold:

$$\forall x.\, (\alpha x + t \not\doteq 0 \vee \phi) \ \equiv\ (\alpha \nmid t) \vee [x/(t \div \alpha)]\phi, \qquad (\alpha \mid t) \ \equiv\ \alpha(t \div \alpha) \doteq t$$

where $\div$ denotes integer division. Vice versa, an expression $c \doteq t \div \alpha$ can be encoded in PAID using axioms like $\alpha c \leq t \ \wedge\ (t < \alpha c + \alpha \vee t < \alpha c - \alpha)$.

# 4 Interpolation for Uninterpreted Functions

## 4.1 A Relational Encoding of Uninterpreted Functions

For practical verification and interpolation problems, uninterpreted functions are more common and often more important than uninterpreted predicates. In the context of interpolation, functions share many properties with predicates; in particular, the quantifier-free fragment PA+UF is again not closed under interpolation, in analogy to Theorem 4.

Similar to the previous section, the interpolation property can be restored by adding means of integer division. To this end, we define the logic PAID+UF like PAID, but allowing arbitrary occurrences of uninterpreted functions in terms. For reasoning and interpolation purposes, we represent functions via an encoding into uninterpreted predicates. The resulting calculus strongly resembles the congruence closure approach used in SMT solvers (e.g., [5]). To formalise the encoding, we introduce a further logic, PAID+UF$_p$. Recall that $P$ and $F$ denote the vocabularies of uninterpreted predicates and functions. We assume that a fresh $(n+1)$-ary uninterpreted predicate $f_p \in P$ exists for every $n$-ary uninterpreted function $f \in F$. The logic PAID+UF$_p$ is then derived from PAID by incorporating occurrences of predicates $f_p$ of the following form:

$$\exists x.\ \big(f_p(t_1, \ldots, t_n, x) \wedge \phi\big) \tag{3}$$

where $x \in X$ ranges over variables, $t_1, \ldots, t_n$ over terms that do not contain $x$, and $\phi$ over PAID+UF$_p$ formulae (possibly containing $x$). In order to avoid universal quantifiers, we do not allow expressions (3) underneath negations.

Formulae in PAID+UF can uniformly be mapped to PAID+UF$_p$ by rewriting:

$$\phi[f(t_1, \ldots, t_n)] \quad \rightsquigarrow \quad \exists x.\ (f_p(t_1, \ldots, t_n, x) \wedge \phi[x]) \tag{4}$$

provided that the terms $t_1, \ldots, t_n$ do not contain variables bound in $\phi$. To stay within PAID+UF$_p$, application of the rule underneath negations has to be avoided, which can be done by transformation to negation normal form. We write $\phi_{Rel}$ for the function-free PAID+UF$_p$ formula derived from a PAID+UF formula $\phi$ by exhaustive application of (4). Vice versa, $\phi$ can be obtained from $\phi_{Rel}$ by applying (4) in the opposite direction. Assuming *functional consistency*, the formulae $\phi$ and $\phi_{Rel}$ are satisfiability-equivalent:

**Lemma 9.** *Let $FC_f$ denote the functional consistency axiom:*[3]

$$FC_f = \quad \forall \bar{x}_1, \bar{x}_2, y_1, y_2.\ \big((f_p(\bar{x}_1, y_1) \wedge f_p(\bar{x}_2, y_2) \wedge \bar{x}_1 \doteq \bar{x}_2) \ \rightarrow\ y_1 \doteq y_2\big) \tag{5}$$

*A PAID+UF formula $\phi$ is satisfiable exactly if $\phi_{Rel} \wedge \bigwedge_{f \in F} FC_f$ is satisfiable.*

By the lemma, it is sufficient to construct a proof of $\neg(\phi_{Rel} \wedge \bigwedge_{f \in F} FC_f)$ in order to show that $\phi$ is unsatisfiable.[4] The axioms $FC_f$ can be handled by ground instantiation, just like the predicate consistency axiom (1): whenever atoms $f_p(\bar{s}_1, t_1)$ and $f_p(\bar{s}_2, t_2)$ occur in the antecedent of a sequent, an instance of $FC_f$ can be generated using the rules ALL-LEFT-L/R and the substitution $[\bar{x}_1/\bar{s}_1, \bar{x}_2/\bar{s}_2, y_1/t_1, y_2/t_2]$. This form of instantiation is sufficient, because predicates $f_p$ only occur in positive positions in $\phi_{Rel}$, and therefore only turn up in antecedents. As before, the number of required instances can be kept under control by formulating suitable refinements [7].

### 4.2 Interpolation for PAID+UF

PAID+UF conjunctions $A \wedge B$ can be interpolated by constructing a proof of

$$\lfloor A_{Rel} \rfloor_L, \lfloor B_{Rel} \rfloor_R, \{\lfloor FC_f \rfloor_L\}_{f \in F_A}, \{\lfloor FC_f \rfloor_R\}_{f \in F_B} \ \vdash\ \emptyset\ \blacktriangleright I \tag{6}$$

where $F_A/F_B$ are the uninterpreted functions occurring in $A/B$. Due to the soundness of the calculus, the existence of a proof guarantees that $I$ is an interpolant. Vice versa, a completeness result corresponding to Thm. 5 also holds for PAID+UF$_p$. Because PAID+UF$_p$ interpolants can be translated back to PAID+UF by virtue of (4), we also have a closure result:

[3] Axiom (5) can also be formulated as $\forall \bar{x}_1, y_1, y_2.\ (f_p(x, y_1) \wedge f_p(\bar{x}, y_2) \rightarrow y_1 \doteq y_2)$, assuming the predicate consistency axiom (1). We chose (5) to avoid having to consider the auxiliary axiom (1) at this point, which simplifies presentation.

[4] Note that this formulation fails to work if arbitrary quantifiers are allowed in $\phi$; this case would require axioms for totality of functions as well.

$$
\dfrac{\cdots \qquad \dfrac{*}{\ldots, \lfloor b \doteq x_1 \rfloor_L, \lfloor x_2 \doteq c \rfloor_L, \lfloor x_1 \doteq x_2 \rfloor_L \ \vdash\ \lfloor b \doteq c \rfloor_R \ \blacktriangleright\ b \doteq c} \qquad \dfrac{*}{\ldots, \lfloor y_1 \doteq y_2 \rfloor_R, \lfloor d \doteq 1 \rfloor_L, \lfloor y_1 \doteq y_2 + d \rfloor_R \ \vdash\ \emptyset \ \blacktriangleright\ d \doteq 1}}{\dfrac{\ldots, \lfloor (f_p(b, y_1) \land f_p(c, y_2) \land b \doteq c) \ \to\ y_1 \doteq y_2 \rfloor_R \ \vdash\ \emptyset \ \blacktriangleright\ b \doteq c \land d \doteq 1}{\ldots, \lfloor f_p(b, y_1) \rfloor_R, \lfloor f_p(c, y_2) \rfloor_R, \lfloor FC_f \rfloor_R, \lfloor x_1 \doteq x_2 \rfloor_L \ \vdash\ \emptyset \ \blacktriangleright\ b \doteq c \land d \doteq 1} \ \text{(ii)}}
$$

$$
\mathcal{D}
$$

$$
\dfrac{\dfrac{\dfrac{\cdots \qquad \dfrac{*}{\ldots, \lfloor a \doteq 1 \rfloor_R \ \vdash\ \lfloor 2 \doteq a + 1 \rfloor_L \ \blacktriangleright\ a \not\doteq 1} \qquad \mathcal{D}}{\ldots, \lfloor (f_p(2, x_1) \land f_p(a+1, x_2) \land 2 \doteq a + 1) \ \to\ x_1 \doteq x_2 \rfloor_L \ \vdash\ \emptyset \ \blacktriangleright\ I_1} \ \textsc{or-left-l}^+}{\ldots, \lfloor f_p(2, x_1) \rfloor_L, \lfloor f_p(a+1, x_2) \rfloor_L, \lfloor FC_f \rfloor_L \ \vdash\ \emptyset \ \blacktriangleright\ I_1} \ \text{(i)}}{\vdots}
$$

$$
\dfrac{\vdots}{\lfloor A_{Rel} \rfloor_L, \lfloor B_{Rel} \rfloor_R, \lfloor FC_f \rfloor_L, \lfloor FC_f \rfloor_R \ \vdash\ \emptyset \ \blacktriangleright\ I_1} \ \textsc{and-left}^+, \textsc{ex-left}^+
$$

**Fig. 3.** Interpolating proof of Example 11. Parts of the proof concerned with arithmetic reasoning or application of the CLOSE-* rules are not shown.

**Theorem 10.** *The logic PAID+UF is closed under interpolation.*

*Example 11.* We consider the PAID+UF interpolation problem $A \land B$ with

$$
A = \ b \doteq f(2) \land f(a+1) \doteq c \land d \doteq 1, \qquad B = \ a \doteq 1 \land f(b) \doteq f(c) + d \ .
$$

The corresponding PAID+UF$_p$ formulae are:

$$
\begin{aligned}
A_{Rel} &= \ \exists x_1.\ \big(f_p(2, x_1) \land \exists x_2.\ \big(f_p(a+1, x_2) \land b \doteq x_1 \land x_2 \doteq c \land d \doteq 1\big)\big) \\
B_{Rel} &= \ \exists y_1.\ \big(f_p(b, y_1) \land \exists y_2.\ \big(f_p(c, y_2) \land a \doteq 1 \land y_1 \doteq y_2 + d\big)\big) \ .
\end{aligned}
$$

The unsatisfiability of $A_{Rel} \land B_{Rel}$ is proven in Fig. 3, requiring two applications of $FC_f$: (i) for the pair $f(2), f(a+1)$, and (ii) for $f(b), f(c)$. The resulting interpolant is $I_1 = \ a \not\doteq 1 \lor (b \doteq c \land d \doteq 1)$ and contains a disjunction due to splitting over an $L$-formula (i), and a conjunction due to (ii). □

As in Lem. 6, a sufficient condition for PAID+UF$_p$ interpolants can be given by restricting applications of the functional consistency axiom:

**Lemma 12.** *Suppose that every instantiation of an axiom $FC_f$ in a proof $\mathcal{P}$ of* (6) *has the form*

$$
\dfrac{\dfrac{\dfrac{\vdots}{\ldots \ \vdash\ \lfloor \bar{s}_1 \doteq \bar{s}_2 \rfloor_F, \ldots \ \blacktriangleright\ J_3}}{\mathcal{Q}} \qquad \dfrac{\dfrac{\vdots}{\ldots, \lfloor t_1 \doteq t_2 \rfloor_F \ \vdash\ \ldots \ \blacktriangleright\ J_4}}{\mathcal{R}}}{}
$$

$$
\dfrac{\dfrac{\dfrac{*}{\lfloor f_p(\bar{s}_1, t_1) \rfloor_D \ \vdash\ \lfloor f_p(\bar{s}_1, t_1) \rfloor_F \ \blacktriangleright\ J_1} \qquad \dfrac{*}{\lfloor f_p(\bar{s}_2, t_2) \rfloor_E \ \vdash\ \lfloor f_p(\bar{s}_2, t_2) \rfloor_F \ \blacktriangleright\ J_2} \qquad \mathcal{Q} \qquad \mathcal{R}}{\ldots, \lfloor (f_p(\bar{s}_1, t_1) \ \land\ f_p(\bar{s}_2, t_2) \ \land\ \bar{s}_1 \doteq \bar{s}_2) \ \to\ t_1 \doteq t_2 \rfloor_F \ \vdash\ \ldots \ \blacktriangleright\ J_5}}{\ldots, \lfloor f_p(\bar{s}_1, t_1) \rfloor_D, \lfloor f_p(\bar{s}_2, t_2) \rfloor_E \ \vdash\ \ldots \ \blacktriangleright\ J_6} \ \textsc{all-left}^+
$$

*where (i) $D, E \in \{L, R\}$ and $F \in \{D, E\}$ are arbitrary labels, (ii) $R \in \{D, E\}$ implies $F = R$, (iii) the proof $\mathcal{Q}$ only uses the rules* RED-RIGHT, MUL-RIGHT, IPI-RIGHT, AND-RIGHT-L, *and* CLOSE-EQ-RIGHT *applied to an equality derived from*

$\bar{s}_1 \doteq \bar{s}_2$ *(see [3] or Appendix A), (iv)* ALL-LEFT *and* EX-RIGHT *are not applied in any other places in $\mathcal{P}$. Then $I$ is a PAID+UF$_p$ formula up to normalisation of guards.*

Proofs of this shape closely correspond to the reasoning of congruence closure procedures (e.g., [5]): two terms/nodes $f(\bar{s}_1)$ and $f(\bar{s}_2)$ are collapsed only once the equations $\bar{s}_1 \doteq \bar{s}_2$ have been derived. Congruence closure can therefore be used to efficiently generate proofs satisfying the conditions of the lemma (abstracting from the additional reasoning necessary to handle the integers).

As in Sect. 3.2, it is also possible to relax the conditions of the lemma; in particular, there is no need to restrict $FC_f$ applications with $D = E$. The resulting interpolation procedure is very flexible, in the sense that many different interpolants can be generated from essentially the same proof. Reordering $FC_f$ applications, for instance, changes the propositional structure of interpolants:

*Example 13.* In Example 11, the interpolant $I_1 = \; a \not\doteq 1 \vee (b \doteq c \wedge d \doteq 1)$ is derived using two $FC_f$ applications (i) and (ii). Reordering the applications, so as to perform (ii) before (i), yields the interpolant $I_2 = \; (a \not\doteq 1 \vee b \doteq c) \wedge d \doteq 1$. □

### 4.3 Interpolation for the Theory of Extensional Arrays

The first-order theory of arrays [9] is typically encoded using uninterpreted function symbols *select* and *store* by means of the following axioms:

$$\forall x, y, z.\ \mathit{select}(\mathit{store}(x, y, z), y) \doteq z \tag{7}$$

$$\forall x, y_1, y_2, z.\ \big(y_1 \doteq y_2 \vee \mathit{select}(\mathit{store}(x, y_1, z), y_2) \doteq \mathit{select}(x, y_2)\big) \tag{8}$$

Intuitively, $\mathit{select}(x, y)$ retrieves the element of array $x$ stored at position $y$, while $\mathit{store}(x, y, z)$ denotes the array that is identical to $x$, except that position $y$ stores value $z$. The *extensional* theory of arrays additionally supports equalities between arrays and is encoded using the following axiom:

$$\forall x_1, x_2.\ (x_1 \doteq x_2 \leftrightarrow (\forall y.\ \mathit{select}(x_1, y) \doteq \mathit{select}(x_2, y))) \tag{9}$$

The quantifier-free theory of arrays is again not closed under interpolation, even without arithmetic, as was already noted in [10, 11]. A classical example is given by the following inconsistent formulae:

$$A = \quad M' \doteq \mathit{store}(M, a, d)$$

$$B = \quad b \not\doteq c \;\wedge\; \mathit{select}(M', b) \not\doteq \mathit{select}(M, b) \;\wedge\; \mathit{select}(M', c) \not\doteq \mathit{select}(M, c)\,,$$

which only permit quantified interpolants, of the form

$$\forall y_1, y_2.\ \big(y_1 \doteq y_2 \;\vee\; \mathit{select}(M, y_1) \doteq \mathit{select}(M', y_1) \;\vee\; \mathit{select}(M, y_2) \doteq \mathit{select}(M', y_2)\big).$$

Naturally, combining array theory with quantifier-free Presburger arithmetic only exacerbates the problem. As we have shown in previous sections, extending PA+UP by guarded integer divisibility predicates results in a theory that is closed under interpolation. We can extend this solution to the theory of arrays, but still only obtain closure under interpolation for small fragments of the logic (like for formulae that do not contain the *store* symbol). The resulting interpolation procedure is similar in flavour to the procedures in [12, 13] and works by explicit instantiation of the array axioms. As in Sect. 3, axioms are handled lazily using the rules ALL-LEFT-L/R, which introduce quantifiers in interpolants as needed.

*Array interpolation via relational encoding.* To reduce array expressions to expressions involving uninterpreted predicates, we use the same relational encoding as in Sect. 4. We first lift the axioms (7), (8), and (9) to the relational encoding:

$$AR_1 = \forall x_1, x_2, y, z_1, z_2.\ \big(\mathit{store}_p(x_1, y, z_1, x_2) \wedge \mathit{select}_p(x_2, y, z_2)\ \to\ z_1 \doteq z_2\big)$$

$$AR_2 = \forall x_1, x_2, y_1, y_2, z, z_1, z_2.\ \left(\begin{array}{l} \quad \mathit{store}_p(x_1, y_1, z, x_2) \\ \wedge\ \mathit{select}_p(x_1, y_2, z_1) \quad \to\ y_1 \doteq y_2 \vee z_1 \doteq z_2 \\ \wedge\ \mathit{select}_p(x_2, y_2, z_2) \end{array}\right)$$

$$AR_3 = \forall x_1, x_2.\ \left(\begin{array}{c} \forall y, z_1, z_2.\ (\mathit{select}_p(x_1, y, z_1) \wedge \mathit{select}_p(x_2, y, z_2) \to z_1 \doteq z_2) \\ \to x_1 \doteq x_2 \end{array}\right)$$

As in the previous sections, these axioms can be used in proofs by ground instantiation based on literals that occur in antecedents of sequents; in the case of $AR_3$, it is also necessary to perform instantiation based on equations occurring in the succedent. This yields an interpolating (though incomplete) calculus for the full logic QPA+AR, and an interpolating decision procedure for the combined theory PAID+AR of Presburger arithmetic with integer division and arrays. Interpolants expressed via the relational encodings of the functions *select* and *store* can be translated into interpolants over array expressions via re-substitution rules.

*Array properties.* The *array property fragment*, introduced by Bradley et al. [14], comprises Presburger arithmetic and the theory of extensional arrays parameterised by suitable element theories. In array property formulae, integer variables may be quantified universally, provided that the matrix of the resulting quantified formula is *guarded* by a Boolean combination of equalities and non-strict inequalities. Using such formulae, one can express properties like equality and sortedness of arrays, as they commonly occur in formulae extracted from programs. Despite its expressiveness, satisfiability for this fragment was shown to be decidable by providing an effective decision procedure [14].

Although Bradley et al. did not consider interpolation for the theory of array properties, we observe that the decision procedure given in [14] can easily be made interpolating using the calculus for QPA+AR provided in this paper. The decision procedure proceeds by reducing, in a sequence of 5 steps, array property formulae to formulae in the combined theory of Presburger arithmetic with uninterpreted functions and the element theories. These 5 steps essentially correspond to instantiation of the array axioms and of quantified parts of the input formulae, which can be implemented using the interpolating rules provided in Fig. 1. The final step is a call to an interpolating decision procedure for Presburger arithmetic and uninterpreted functions combined with suitable element theories; we have presented such a procedure in this paper.

We remark that the array property fragment is not subsumed by the restriction of QPA+AR to Presburger arithmetic and array theory with guarded quantification as allowed in PAID+UF.

## 5 Related Work and Conclusion

**Related work.** Yorsh et al. [15] present a combination method to generate interpolants using interpolation procedures for individual theories. To be applicable, the method requires individual theories to be *equality interpolating*; this is neither the case for Presburger arithmetic nor for arrays. To the best of our knowledge, it is unknown whether quantifier-free Presburger arithmetic with the integer division operator $\div$ is equality interpolating.

Interpolation procedures for uninterpreted functions are given by McMillan [10] and Fuchs et al. [16]. The former approach uses an interpolating calculus with rules for transitivity, congruence, etc.; the latter is based on congruence closure algorithms. Our calculus in Sect. 4 has similarities with [16], but is more flexible concerning the order in which congruence rules are applied. A more systematic comparison is planned as future work, including estimating the cost of interpolating uninterpreted functions via a reduction to predicates, rather than via some direct procedure. The papers [10, 16] do not consider the combination with full Presburger arithmetic.

Kapur et al. [11] present an interpolation method for arrays that works by reduction to the theory of uninterpreted functions. To some degree, the interpolation procedure of Sect. 4.3 can be considered as a lazy version of the procedure in [11], performing the reduction to uninterpreted functions only on demand.

In [12], Jhala et al. define a *split prover* that computes quantifier-free interpolants in a fragment of the theory of arrays, among others. The main objective of [12] is to derive interpolants in restricted languages, which makes it possible to guarantee convergence and a certain form of completeness in model checking. While our procedure is more general in that the full combined theory of PA with arrays can be handled, we consider it as important future work to integrate techniques to restrict interpolant languages into our procedure.

McMillan provides a complete procedure to generate (potentially) quantified interpolants for the full theory of arrays [13] by means of explicit array axioms. Our interpolation method resembles McMillan's in that explicit array axioms are given to a theorem prover, but our procedure is also complete in combination with Presburger arithmetic.

Bradley et al. introduce the concept of constrained universal quantification in array theory [14], which essentially allows a single universal array index quantifier, possibly restricted to an index subrange, e.g. all indices in some range $[l, u]$. Unlike full quantified array theory, satisfiability is decidable in Bradley's fragment; interpolation is not considered in this work. We have discussed the relationship of this fragment to QPA+AR in Section 4.3.

For a discussion of related work concerning interpolation in pure quantifier-free Presburger arithmetic, we refer the reader to [3].

**Conclusion.** We have presented interpolating calculi for the theories of Presburger arithmetic combined with uninterpreted predicates (QPA+UP), uninterpreted functions (QPA+UF), and extensional arrays (QPA+AR). We have demonstrated that these extensions require the use of quantifiers in interpolants. Adding notions of *guarded quantification*, we therefore identified fragments of the full first-order theories that are closed under interpolation, yet are expressible in assertion languages present in standard programming languages.

As future work, we plan to extend our results to interpolating SMT solvers, particularly aiming at procedures that can be used in model checkers based on the *lazy abstraction with interpolants* paradigm. On the theoretical side, we will study the relationship between the logics discussed in this paper, and architectures for combining interpolating procedures, e.g., [15]. We also plan to investigate, possibly along the lines of [17], how our interpolation procedure for uninterpreted functions relates to existing methods [10, 16], and how it affects the strength of computed interpolants. Finally, we plan to investigate a combination of our calculus with the Split-Prover approach in [12].

# A List of interpolating sequent rules

$$\frac{\begin{array}{c}\Gamma, \lfloor\phi\rfloor_L \vdash \Delta \blacktriangleright I \\ \Gamma, \lfloor\psi\rfloor_L \vdash \Delta \blacktriangleright J\end{array}}{\Gamma, \lfloor\phi \vee \psi\rfloor_L \vdash \Delta \blacktriangleright I \vee J}\;\textsc{or-left-l} \qquad \frac{\begin{array}{c}\Gamma, \lfloor\phi\rfloor_R \vdash \Delta \blacktriangleright I \\ \Gamma, \lfloor\psi\rfloor_R \vdash \Delta \blacktriangleright J\end{array}}{\Gamma, \lfloor\phi \vee \psi\rfloor_R \vdash \Delta \blacktriangleright I \wedge J}\;\textsc{or-left-r}$$

$$\frac{\begin{array}{c}\Gamma \vdash \lfloor\phi\rfloor_L, \Delta \blacktriangleright I \\ \Gamma \vdash \lfloor\psi\rfloor_L, \Delta \blacktriangleright J\end{array}}{\Gamma \vdash \lfloor\phi \wedge \psi\rfloor_L, \Delta \blacktriangleright I \vee J}\;\textsc{and-right-l} \qquad \frac{\begin{array}{c}\Gamma \vdash \lfloor\phi\rfloor_R, \Delta \blacktriangleright I \\ \Gamma \vdash \lfloor\psi\rfloor_R, \Delta \blacktriangleright J\end{array}}{\Gamma \vdash \lfloor\phi \wedge \psi\rfloor_R, \Delta \blacktriangleright I \wedge J}\;\textsc{and-right-r}$$

$$\frac{\Gamma, \lfloor\phi\rfloor_D, \lfloor\psi\rfloor_D \vdash \Delta \blacktriangleright I}{\Gamma, \lfloor\phi \wedge \psi\rfloor_D \vdash \Delta \blacktriangleright I}\;\textsc{and-left} \qquad \frac{\Gamma \vdash \lfloor\phi\rfloor_D, \lfloor\psi\rfloor_D, \Delta \blacktriangleright I}{\Gamma \vdash \lfloor\phi \vee \psi\rfloor_D, \Delta \blacktriangleright I}\;\textsc{or-right}$$

$$\frac{\Gamma \vdash \lfloor\phi\rfloor_D, \Delta \blacktriangleright I}{\Gamma, \lfloor\neg\phi\rfloor_D \vdash \Delta \blacktriangleright I}\;\textsc{not-left} \qquad \frac{\Gamma, \lfloor\phi\rfloor_D \vdash \Delta \blacktriangleright I}{\Gamma \vdash \lfloor\neg\phi\rfloor_D, \Delta \blacktriangleright I}\;\textsc{not-right}$$

$$\frac{*}{\Gamma, \lfloor\phi\rfloor_L \vdash \lfloor\phi\rfloor_L, \Delta \blacktriangleright \mathit{false}}\;\textsc{close-ll} \qquad \frac{*}{\Gamma, \lfloor\phi\rfloor_R \vdash \lfloor\phi\rfloor_R, \Delta \blacktriangleright \mathit{true}}\;\textsc{close-rr}$$

$$\frac{*}{\Gamma, \lfloor\phi\rfloor_L \vdash \lfloor\phi\rfloor_R, \Delta \blacktriangleright \phi}\;\textsc{close-lr} \qquad \frac{*}{\Gamma, \lfloor\phi\rfloor_R \vdash \lfloor\phi\rfloor_L, \Delta \blacktriangleright \neg\phi}\;\textsc{close-rl}$$

$$\frac{\Gamma, \lfloor[x/t]\phi\rfloor_L, \lfloor\forall x.\phi\rfloor_L \vdash \Delta \blacktriangleright I}{\Gamma, \lfloor\forall x.\phi\rfloor_L \vdash \Delta \blacktriangleright \forall_{Rt}\, I}\;\textsc{all-left-l} \qquad \frac{\Gamma, \lfloor[x/t]\phi\rfloor_R, \lfloor\forall x.\phi\rfloor_R \vdash \Delta \blacktriangleright I}{\Gamma, \lfloor\forall x.\phi\rfloor_R \vdash \Delta \blacktriangleright \exists_{Lt}\, I}\;\textsc{all-left-r}$$

$$\frac{\Gamma, \lfloor[x/c]\phi\rfloor_D \vdash \Delta \blacktriangleright I}{\Gamma, \lfloor\exists x.\phi\rfloor_D \vdash \Delta \blacktriangleright I}\;\textsc{ex-left} \qquad \frac{\Gamma \vdash \lfloor[x/c]\phi\rfloor_D, \Delta \blacktriangleright I}{\Gamma \vdash \lfloor\forall x.\phi\rfloor_D, \Delta \blacktriangleright I}\;\textsc{all-right}$$

**Fig. 4.** The upper frame presents all interpolating rules for propositional logic, while the lower frame shows the interpolating rules to handle quantifiers. Parameter $D$ stands for either $L$ or $R$. The quantifier $\forall_{Rt}$ denotes universal quantification over all constants occurring in $t$ but not in $\Gamma_L \cup \Delta_L$; likewise, $\exists_{Lt}$ denotes existential quantification over all constants occurring in $t$ but not in $\Gamma_R \cup \Delta_R$.

$$\frac{\Gamma, \lfloor (\alpha \nmid t) \vee \exists x.\ (\alpha x + t \doteq 0 \wedge \phi) \rfloor_D \ \vdash \ \Delta \ \blacktriangleright I}{\Gamma, \lfloor \forall x.\ (\alpha x + t \not\doteq 0 \vee \phi) \rfloor_D \ \vdash \ \Delta \ \blacktriangleright I} \ \text{ALL-LEFT-GRD}$$

$$\frac{\Gamma \ \vdash \ \lfloor (\alpha \mid t) \wedge \forall x.\ (\alpha x + t \not\doteq 0 \vee \phi) \rfloor_D, \Delta \ \blacktriangleright I}{\Gamma \ \vdash \ \lfloor \exists x.\ (\alpha x + t \doteq 0 \wedge \phi) \rfloor_D, \Delta \ \blacktriangleright I} \ \text{EX-RIGHT-GRD}$$

$$\frac{*}{\Gamma \ \vdash \ 0 \doteq 0 \, [t^A \doteq 0], \Delta \ \blacktriangleright \exists_{LA} \, t^A \not\doteq 0} \ \text{CLOSE-EQ-RIGHT}$$

$$\frac{\Gamma \ \vdash \ t \doteq 0 \, [0 \not\doteq 0], \lfloor t \doteq 0 \rfloor_R, \Delta \ \blacktriangleright I}{\Gamma \ \vdash \ \lfloor t \doteq 0 \rfloor_R, \Delta \ \blacktriangleright I} \ \text{IPI-RIGHT}$$

$$\frac{\Gamma, t \doteq 0 \, [t^A \doteq 0] \ \vdash \ s + \alpha \cdot t \doteq 0 \, [s^A + \alpha \cdot t^A \circ 0], \Delta \ \blacktriangleright I}{\Gamma, t \doteq 0 \, [t^A \doteq 0] \ \vdash \ s \doteq 0 \, [s^A \circ 0], \Delta \ \blacktriangleright I} \ \text{RED-RIGHT}$$

$$\frac{\Gamma \ \vdash \ \alpha \cdot t \doteq 0 \, [\alpha \cdot t^A \circ 0], \Delta \ \blacktriangleright I}{\Gamma \ \vdash \ t \doteq 0 \, [t^A \circ 0], \Delta \ \blacktriangleright I} \ \text{MUL-RIGHT}$$

**Fig. 5.** The lower frame presents an excerpt of rules for PA, while the upper frame shows the additional rules for the PAID+UP extension. Parameter $D$ stands for either $L$ or $R$. In CLOSE-*, $\exists_{LA}$ denotes existential quantification $\exists c_1, \dots, c_n.$, where $c_1, \dots, c_n$ are the constants that occur in $\Gamma_L, \Delta_L$ but not in $\Gamma_R, \Delta_R$. In RED-RIGHT and MUL-RIGHT, $\circ \in \{\doteq, \not\doteq\}$. In MUL-RIGHT, $\alpha > 0$ is a positive literal. Formulae in squared brackets such as $[t^A \doteq 0]$ denote *partial interpolants*, which are required for rules mixing left and right parts. We refer to [3] for more details on partial interpolants.

## B Theorem 4: PA+UP is not Closed under Interpolation

This section proves Theorem 4, using of the following intermediate result:

**Lemma 14.** *Let $y$ be a constant and $S = \{\alpha_i y + \beta_i \mid \alpha_i, \beta_i \in \mathbb{Z}, i \in \{1, \ldots, n\}\}$ be a finite set of terms in PA. Then there exists an even number $a \in 2\mathbb{Z}$ such that $\frac{a}{2} \notin \{val_{y \mapsto a}(t) \mid t \in S\}$.*

*Proof.* Choose $a \in 2\mathbb{Z}$ such that $a > 2 \cdot \max_i |\beta_i|$. Let us suppose that, for some $t = \alpha y + \beta \in S$, we have $val_{y \mapsto a}(\alpha y + \beta) = \alpha a + \beta = \frac{a}{2}$. Thus $2\alpha a + 2\beta = a$ and therefore $(2\alpha - 1)a = -2\beta$. Since $2\alpha - 1 \neq 0$, we distinguish two cases:

- $2\alpha - 1 > 0$: this yields a contradiction because $(2\alpha - 1)a \geq a > 2 \cdot |\beta| = |-2\beta| \geq -2\beta$.
- $2\alpha - 1 < 0$: this yields a contradiction because $(2\alpha - 1)a \leq -a < -2 \cdot |\beta| = -|2\beta| \leq -2\beta$. □

We can now prove Theorem 4.

*Proof (Proof of Theorem 4).* We construct an example of inconsistent formulae $A$ and $B$ in PA+UP whose interpolant requires quantification. Consider:

$$A = \quad 2c - y \doteq 0 \wedge p(c) \qquad\qquad B = \quad 2d - y \doteq 0 \wedge \neg p(d)$$

The symbols $p$ and $y$ are common, while $c$ and $d$ are local. The conjunction $A \wedge B$ is unsatisfiable. The strongest and the weakest interpolants for $A$ and $B$ are, respectively:

$$I_s = \quad \exists x.\, (2x - y \doteq 0 \wedge p(x)) \qquad I_w = \quad \forall x.\, (2x - y \doteq 0 \rightarrow p(x))$$

Now suppose $I$ is a quantifier-free interpolant for $A \wedge B$; in particular, $I$ contains only the common symbols $p$ and $y$. Let $S = \{t \mid p(t) \text{ occurs in } I\}$ be the set of all terms occurring in $I$ as arguments of $p$. All elements of $S$ are PA terms over the symbol $y$. By Lem. 14, there is an even number $a \in 2\mathbb{Z}$ such that $\frac{a}{2} \notin \{val_{y \mapsto a}(t) \mid t \in S\}$.

Since $I$ is an interpolant, the implications $I_s \Rightarrow I$ and $I \Rightarrow I_w$ hold. In particular, observe that

$$(2 \mid y) \models (I_s \leftrightarrow I) \wedge (I \leftrightarrow I_w)\,. \qquad (10)$$

Choose an interpretation $K$ with $K(y) = a$ that satisfies $I$ (this is possible, because such satisfying interpretations exist for $I_s$). Because of (10) and because $K(y)$ is even, it holds that $\frac{K(y)}{2} \in K(p)$. However, we know that $I$ does not contain any atom $p(t)$ such that $val_K(t) = \frac{K(y)}{2}$. This means that $I$ is also satisfied by the interpretation $K'$ that coincides with $K$, with the only exception that $\frac{K'(y)}{2} \notin K'(p)$. But $K'$ violates $I_w$, contradicting the assumption that $I$ is an interpolant. □

## C Proofs generating PAID+UP Interpolants

We give a proof of Lem. 6, from which Thm. 5 can be derived by providing a PAID+UP proof procedure. In the whole section, we use rules introduced in [3].

### C.1 Proof of Lemma 6: Sufficient Conditions for PAID+UP Interpolants

The only rules that introduce quantifiers in interpolants in $\mathcal{P}$ are (i) the rules CLOSE-EQ-*, CLOSE-INEQ, and (ii) the rules ALL-LEFT-* that are used to instantiate axiom (1). The quantifiers generated by the first kind of rules can directly be eliminated, because the body of the quantified expression is an arithmetic literal. In the second case, we consider the sub-proof $\mathcal{Q}$, as described in the lemma. There are different scenarios depending on the values of $D, E, F$; for sake of presentation, we only consider $D = L, E = R, F = L$ (all other cases are similar):

$$
\dfrac{\cdots \qquad \dfrac{\dfrac{\dfrac{*}{\ldots \vdash 0 \doteq 0\,[u_i \doteq 0], \ldots \ \blacktriangleright K_i}\ \textsc{close-eq-right}}{\ldots \vdash s_i - t_i \doteq 0\,[s_i - t_i \doteq 0], \ldots \ \blacktriangleright K_i}\ \textsc{red-right}^*, \textsc{mul-right}^*}{\ldots \vdash \lfloor s_i - t_i \doteq 0 \rfloor_L, \ldots \ \blacktriangleright K_i}\ \textsc{ipi-right} \qquad \cdots}{\dfrac{\Gamma', \lfloor p(\bar{s}) \rfloor_L \vdash \lfloor \bar{s} - \bar{t} \doteq 0 \rfloor_L, \lfloor p(\bar{t}) \rfloor_R, \Delta' \ \blacktriangleright \bigvee_i K_i}{\mathcal{Q}}}\ \textsc{and-right}^*
$$

It is possible that $\mathcal{Q}$ contains further applications of RED-RIGHT, MUL-RIGHT, IPI-RIGHT, or AND-RIGHT-L in between the steps shown, but this has no effect on the shape of the interpolant $\bigvee_i K_i$ (apart from some disjunct $K_i$ possibly occurring multiple times). The rule CLOSE-EQ-RIGHT generates the interpolant $K_i = (\exists_{LA}\ u_i \not\doteq 0)$. A careful analysis of the calculus shows that the quantifier $\exists_{LA}$ is in fact empty, i.e., $K_i = (u_i \not\doteq 0)$ and $J_2 = (\bigvee_i K_i) = (\bigvee_i u_i \not\doteq 0)$.

We need to analyse the shape of the interpolant

$$
\begin{aligned}
J_5 \ &= \ \forall_{R\bar{s}\bar{t}}\ J_4 \ = \ \forall_{R\bar{t}}\ J_4 \ = \ \forall_{R\bar{t}}\ \big(J_2 \vee p(\bar{t})\big) \\
&= \ \forall_{R\bar{t}} \left( \bigvee_i u_i \not\doteq 0 \vee p(\bar{t}) \right) \ = \ \forall x_1, \ldots, x_n. \left( \bigvee_i u_i \not\doteq 0 \vee p(\bar{t}) \right)
\end{aligned}
\tag{11}
$$

where $x_1, \ldots, x_n$ are all constants in $\bar{t}$ that are $R$-local in the sequent

$$
\Gamma', \lfloor p(\bar{s}) \rfloor_L \vdash \lfloor \bar{s} - \bar{t} \doteq 0 \rfloor_L, \lfloor p(\bar{t}) \rfloor_R, \Delta' \ .
$$

Using vector notation for $\bar{x} = (x_1, \ldots, x_n)^t$, the atom $p(\bar{t})$ can be represented as $p(\bar{c}_1\bar{x} + v_1, \ldots, \bar{c}_k\bar{x} + v_k)$, where $\bar{c}_1, \ldots, \bar{c}_n \in \mathbb{Z}^n$ are row vectors of coefficients, and $v_1, \ldots, v_k$ are terms that do not contain any of the constants $x_1, \ldots, x_n$. In matrix notation, this gives $p(\bar{t}) = p(C\bar{x} + \bar{v})$ for $C = (\bar{c}_1^t \cdots \bar{c}_k^t)^t \in \mathbb{Z}^{k \times n}$.

Because $x_1, \ldots, x_n$ are $R$-local, we know that these constants do not occur in any partial interpolant in $\Gamma', \lfloor p(\bar{s}) \rfloor_L \vdash \lfloor \bar{s} - \bar{t} \doteq 0 \rfloor_L, \lfloor p(\bar{t}) \rfloor_R, \Delta'$. This implies that the term $-\bar{c}_i\bar{x}$ in the partial interpolant of $s_i - t_i \doteq 0$ $[s_i - t_i \doteq 0]$ will not be affected by any application of RED-RIGHT; likewise, applications of MUL-RIGHT can only introduce scaling by some factor $\alpha$. It is therefore possible to represent the final partial interpolant $u_i \doteq 0$ in the form $\alpha\bar{c}_i\bar{x} + u'_i \doteq 0$, where $\alpha \in \mathbb{Z} \setminus \{0\}$ and $u'_i$ does not contain any of the constants $x_1, \ldots, x_n$. This means that

$$\bigvee_i u_i \not\doteq 0 \quad \equiv \quad \neg \bigwedge_i \alpha\bar{c}_i\bar{x} + u'_i \doteq 0 \quad \equiv \quad \alpha C\bar{x} + \bar{u}' \not\doteq 0$$

We now consider the Smith decomposition [18] of the matrix $C$, i.e., the decomposition $C = LSR$ into three integer matrices, such that (i) $L \in \mathbb{Z}^{k\times k}$ and $R \in \mathbb{Z}^{n\times n}$ are invertible (over integers), (ii) $S$ has the shape

$$\begin{pmatrix} \beta_1 & 0 & \cdots & & & \cdots & 0 \\ 0 & \beta_2 & \ddots & & & & \vdots \\ \vdots & \ddots & \ddots & \ddots & & & \\ & & \ddots & \beta_r & 0 & & \\ \vdots & & & 0 & 0 & & \vdots \\ 0 & \cdots & & & & \cdots & 0 \end{pmatrix}$$

where $r \leq \min\{k, n\}$ and $\beta_1, \ldots, \beta_r$ are positive integers such that $\beta_{i+1} \in \beta_i\mathbb{Z}$ for all $i \in \{1, \ldots, r-1\}$.

The interpolant $J_5$ in (11) can then be rewritten to form (2) as follows:

$$\begin{aligned} J_5 \;&\equiv\; \forall \bar{x}.\ \big(\alpha C\bar{x} + \bar{u}' \not\doteq 0 \;\vee\; p(C\bar{x} + \bar{v})\big) \\ &\equiv\; \forall \bar{x}.\ \big(\alpha LSR\bar{x} + \bar{u}' \not\doteq 0 \;\vee\; p(LSR\bar{x} + \bar{v})\big) \\ &\equiv\; \forall \bar{y}.\ \big(\alpha LS\bar{y} + \bar{u}' \not\doteq 0 \;\vee\; p(LS\bar{y} + \bar{v})\big) \\ &\equiv\; \forall \bar{y}.\ \big(\alpha S\bar{y} + L^{-1}\bar{u}' \not\doteq 0 \;\vee\; p(LS\bar{y} + \bar{v})\big) \\ &\equiv\; \forall y_1.\ \big(\alpha\beta_1 y_1 + (L^{-1}\bar{u}')_1 \not\doteq 0 \;\vee \\ &\phantom{\equiv\;} \forall y_2.\ \big(\alpha\beta_2 y_2 + (L^{-1}\bar{u}')_2 \not\doteq 0 \;\vee \\ &\phantom{\equiv\;} \quad \vdots \\ &\phantom{\equiv\;} \forall y_r.\ \big(\alpha\beta_r y_r + (L^{-1}\bar{u}')_r \not\doteq 0 \;\vee\; p(LS\bar{y} + \bar{v})\big) \cdots \big) \\ &\phantom{\equiv\;} \vee \bigvee_{i=r+1}^{k} (L^{-1}\bar{u}')_i \not\doteq 0 \end{aligned}$$

where $\bar{y} = (y_1, \ldots, y_n)^t$ is a vector of fresh variables, and $(L^{-1}\bar{u}')_i$ denotes the $i$th element of the vector $L^{-1}\bar{u}'$ of terms. Note that the variables $y_{r+1}, \ldots, y_n$ only occur with coefficient zero in the expression $S\bar{y}$, and therefore do not have to be quantified. This shows that $J_5$ is equivalent to a PAID+UP formula and concludes the proof.

## C.2 Proof of Theorem 5: Completeness of the PAID+UP Calculus

We describe a proof procedure that, given a sequent $\Gamma \vdash \Delta \blacktriangleright ?$ such that $\Gamma_L, \Gamma_R \vdash \Delta_L, \Delta_R$ is valid, generates a proof satisfying the conditions in Lem. 6. The following reasoning steps are performed:

1. Apply rules OR-*, AND-*, NOT-*, EX-LEFT, ALL-RIGHT, ALL-LEFT-GRD, EX-RIGHT-GRD, DIV-*, IPI-*, SPLIT-* exhaustively; move all inequalities to the antecedent. This will eliminate all propositional connectives and quantifiers in formulae, what remains are proof goals of the form

$$\begin{array}{c}\{t_i \leq 0\,[t_i^A \leq 0]\}_{i\in I}, \{s_j \doteq 0\,[s_j^A \doteq 0]\}_{j\in J}, \\ \{\lfloor p_k(\bar{u}_k)\rfloor_{D_k}\}_{k\in K}\end{array} \vdash \{\lfloor p_m(\bar{u}_m)\rfloor_{D_m}\}_{m\in M} \blacktriangleright ?$$

where $I, J, K, M$ are disjoint sets of indexes.

2. Apply rules RED-LEFT, COL-RED-*, MUL-LEFT to solve the equalities in the antecedents, as described in [19]. This either leads to an unsatisfiable equality, in which case the rule CLOSE-EQ-LEFT can be applied, or to goals of the form

$$\begin{array}{c}\{t_i \leq 0\,[t_i^A \leq 0]\}_{i\in I}, \\ \{\alpha_j c_j + s_j \doteq 0\,[s_j^A \doteq 0]\}_{j\in J}, \\ \{\lfloor p_k(\bar{u}_k)\rfloor_{D_k}\}_{k\in K}\end{array} \vdash \{\lfloor p_m(\bar{u}_m)\rfloor_{D_m}\}_{m\in M} \blacktriangleright ? \qquad (12)$$

where $I, J, K, M$ are disjoint sets of indexes, $\alpha_j \in \mathbb{Z} \setminus \{0\}$ divides all coefficients and constant terms in $s_j$, the constants $c_j$ are pairwise distinct, and no $c_j$ occurs in any term $t_i$ or $s_{j'}$. In particular, this means that the equalities $\{\alpha_j c_j + s_j \doteq 0\}_{j\in J}$ are satisfiable.

3. Whenever a sequent contains literals $p_k(\bar{u}_k)$ and $p_m(\bar{u}_m)$ such that $p_k = p_m$, and such that $\bar{u}_k \doteq \bar{u}_m$ is implied by the equalities $\{\alpha_j c_j + s_j \doteq 0\}_{j\in J}$, instantiate the consistency axiom (1) for $p_k(\bar{u}_k)$ and $p_m(\bar{u}_m)$, and close the resulting sub-proofs as shown in Lem. 6 (i) and in the beginning of Sect. C.1.
4. Apply STRENGTHEN in a fair manner to the inequalities in the antecedents. Whenever a new equation is generated by a STRENGTHEN application, go back to step 2. Whenever a sequent has been derived in which the inequalities in the antecedent are rationally inconsistent, apply FM-ELIM exhaustively, and apply CLOSE-INEQ to a resulting contradictory inequality.

This procedure will in finitely many steps construct a closed proof tree for the valid PAID+UP sequent $\Gamma \vdash \Delta$; by construction, the proof satisfies the conditions in Lem. 6 (i).

Two steps in the procedure require further considerations:

- *Termination of the loop 2–4:* it has to be shown that systematic application of STRENGTHEN terminates: on every branch of the generated proof, eventually a sequent is reached in which no inequalities remain, or in which the remaining inequalities are rationally inconsistent. Recall that every application of STRENGTHEN produces three new goals: one in which an inequality $t_i \leq 0$ has been turned into an equality $t_i \doteq 0$ (case (a)), and two in which an inequality $t_i \leq 0$ has been strengthened to $t_i + 1 \leq 0$ (case (b)).
Reasoning by contradiction, assume that the procedure never terminates on some branch. This means that, from some point on, we are always looking at the (b) case on the branch, and that the number of inequalities on the branch remains constant and non-zero.
Note that we can assume that each sequent (12) considered in step 4 is valid (ignoring interpolant annotations, which are not relevant at this point); equivalently, the following formula is unsatisfiable:

$$\bigwedge_{i \in I} t_i \leq 0 \ \land \ \bigwedge_{j \in J} \alpha_j c_j + s_j \doteq 0 \ \land \ \bigwedge_{k \in K,\, m \in M} \bar{u}_k \not\doteq \bar{u}_m$$

By rewriting the negated equalities using the equalities $\alpha_j c_j + s_j \doteq 0$, eliminating every occurrence of a constant $c_j$, we obtain a new unsatisfiable conjunction without positive equalities:

$$\bigwedge_{i \in I} t_i \leq 0 \ \land \ \bigwedge_{k \in K,\, m \in M} \bar{u}'_k \not\doteq \bar{u}'_m$$

Because step 3 has not been able to close the goal at hand, we can assume that each disjunction $\bar{u}'_k - \bar{u}'_m \not\doteq 0$ contains at least one equality that is not of the form $0 \not\doteq 0$; we denote this equality with $v_{k,m} \not\doteq 0$. We then know that also the following formula is unsatisfiable:

$$\bigwedge_{i \in I} t_i \leq 0 \ \land \ \bigwedge_{k \in K,\, m \in M} v_{k,m} \not\doteq 0$$

This corresponds to (the negation of) formula (15) in Lem. 15, which tells us that there is an $r \in \mathbb{R}$, and therefore also a $\beta \in \mathbb{Z}$, such that the following formula is even rationally unsatisfiable:

$$\bigwedge_{i \in I} t_i + \beta \leq 0$$

Because fair application of STRENGTHEN will eventually turn every inequality $t_i \leq 0$ into an inequality $t_i + \beta_i \leq 0$ such that $\beta_i \geq \beta$, it is guaranteed that the inequalities in the antecedent eventually become rationally unsatisfiable. This contradicts the assumption that the procedure does not terminate on the considered branch.

– *Existence of a complementary pair in step 3:* we have to show that a complementary pair of literals can be selected in step 3 once all inequalities have been eliminated from a sequent (in step 4). By assumption, we know that the sequents considered in step 3 are valid (again ignoring interpolants). Inequalities-less sequents of the form produced in step 2 are valid iff the sequent

$$\{\alpha_j c_j + s_j \doteq 0\}_{j \in J} \ \vdash \ \{\bar{u}_k \doteq \bar{u}_m\}_{k \in K,\, m \in M} \tag{13}$$

is valid.
We reason by contradiction: suppose that each conjunction $\bar{u}_k \doteq \bar{u}_m$ of equalities contains one equation that is not implied by $\{c_j + s_j/\alpha_j \doteq 0\}_{j \in J}$; we assume w.l.o.g. that this is always the first equation $u_k^1 \doteq u_m^1$. This means that the sequents

$$\{\alpha_j c_j + s_j \doteq 0\}_{j \in J} \;\vdash\; u_k^1 \doteq u_m^1$$

cannot be proven using the rules RED-RIGHT and MUL-RIGHT to reduce equalities in the succedent, and the rule CLOSE-*-RIGHT to detect valid equalities. Consequently, the rules are not sufficient to prove the sequent

$$\{\alpha_j c_j + s_j \doteq 0\}_{j \in J} \;\vdash\; \{u_k^1 \doteq u_m^1\}_{k \in K,\, m \in M}$$

either. By completeness results in [20], this implies that the sequent is invalid. But then also (13) is invalid, contradicting the assumption.

## D Integer Projection Lemma

Let $\{v_1, \ldots, v_n\}$ be a fixed set of variables. For any term $t$, we introduce the function $\bar{t}\colon \mathbb{R}^n \to \mathbb{R}$ defined by $\bar{t}(x_1, \ldots, x_n) = [v_1/x_1, \ldots, v_n/x_n]t$.

**Lemma 15.** *Let $\{t^1, \ldots, t^m\}$ be a set of terms of the form $t^j = c_0^j + \sum_{i=1}^n c_i^j v_i$, and $\{s^1, \ldots, s^p\}$ be a set of non-null terms of the form $s^k = d_0^k + \sum_{i=1}^n d_i^k v_i$, i.e. for each $k$, there exists an $i$ with $d_i^k \neq 0$. Suppose the following formula is valid:*

$$\forall r \in \mathbb{R}\ \exists y_1, \ldots, y_n \in \mathbb{R}\ \forall j \in \{1, \ldots, m\} : \overline{t^j}(y_1, \ldots, y_n) \leq r\,. \tag{14}$$

*Then the following formula is valid:*

$$\begin{aligned} \exists z_1, \ldots, z_n \in \mathbb{Z}\ ( \quad & \forall j \in \{1, \ldots, m\} : \overline{t^j}(z_1, \ldots, z_n) \leq 0 \\ \wedge\ & \forall k \in \{1, \ldots, p\} : \overline{s^k}(z_1, \ldots, z_n) \neq 0\ )\,. \end{aligned} \tag{15}$$

Before we can prove this lemma, we need a few definitions and auxiliary properties. Define the function $f\colon \mathbb{R} \to \mathbb{R}$ via

$$f(x_1, \ldots, x_n) = \max_{j=1}^m \overline{t^j}(x_1, \ldots, x_n)\,.$$

We also define $||C|| = \sum_{i=1}^n \sum_{j=1}^m |c_i^j|$, the 1-norm of the coefficient matrix induced by the $t^j$.

*Property 16.* For real numbers $a$, $b$, $\varepsilon$, $\max\{a+\varepsilon, b\} \leq \max\{a, b\} + |\varepsilon|$.

Proof:

(i) If $\varepsilon \geq 0$, then $a + \varepsilon \leq \max\{a, b\} + \varepsilon$, and $b \leq b + \varepsilon \leq \max\{a, b\} + \varepsilon$. Thus $\max\{a+\varepsilon, b\} \leq \max\{a, b\} + \varepsilon \leq \max\{a, b\} + |\varepsilon|$.
(ii) If $\varepsilon < 0$, then let $\delta = -\varepsilon \geq 0$. Using (i) with $\delta$ in place of $\varepsilon$, we obtain: $\max\{a+\varepsilon, b\} \leq \max\{a+\delta, b\} \leq \max\{a, b\} + \delta = \max\{a, b\} + |\varepsilon|$.

*Property 17.* Given $y_1, \ldots, y_n \in \mathbb{R}$, define $z_i := \lfloor y_i \rfloor \in \mathbb{Z}$ for $i \in \{1, \ldots, n\}$, where $\lfloor\cdot\rfloor$ denotes the floor of a real number. Then $f(z_1, \ldots, z_n) \leq f(y_1, \ldots, y_n) + ||C||$.

Proof:

$$\begin{aligned} f(z_1, \ldots, z_n) &= \max_{j=1}^m \overline{t^j}(z_1, \ldots, z_n) \\ &= \max_{j=1}^m \left( c_0^j + \sum_{i=1}^n c_i^j (y_i + z_i - y_i) \right) \\ &= \max_{j=1}^m \left( c_0^j + \sum_{i=1}^n c_i^j y_i + \sum_{i=1}^n c_i^j (z_i - y_i) \right) \\ &\overset{(*)}{\leq} \underbrace{\max_{j=1}^m \left( c_0^j + \sum_{i=1}^n c_i^j y_i \right)} + \sum_{j=1}^m \left| \sum_{i=1}^n c_i^j (z_i - y_i) \right| \\ &\leq f(y_1, \ldots, y_n) + \sum_{j=1}^m \sum_{i=1}^n |c_i^j| \cdot |z_i - y_i| \\ &\overset{(**)}{\leq} f(y_1, \ldots, y_n) + ||C||\,, \end{aligned}$$

where $(*)$ applies property 16 ($m$ times), and $(**)$ uses $|z_i - y_i| = |\lfloor y_i \rfloor - y_i| \leq 1$.

**Proof of Lemma 15:** Let $r = -(p^2+1) \cdot ||C||$. By (14), there exist $y_1, \ldots, y_n \in \mathbb{R}$ such that $f(y_1, \ldots, y_n) \leq r$. For $1 \leq i \leq n$, define $g_i = \lfloor y_i \rfloor \in \mathbb{Z}$. Our final solutions $z_i$ will have the form $z_i = g_i + h_i$. We obtain suitable $h_i \in \mathbb{Z}$ by examining the condition that the functions $\overline{s^k}$ must not evaluate to 0. The condition involving the $\overline{t_j}$ will then be satisfied due to our choice of $r$ above.

To this end, we prove that there exist integers $h_1, \ldots, h_n \in \mathbb{Z}_{\geq 0}$ such that:

(i) for all $k \in \{1, \ldots, p\}$, $\overline{s^k}(g_1 + h_1, \ldots, g_n + h_n) \neq 0$, and

(ii) for all $i \in \{1, \ldots, n\}$, $h_i \leq p^2$.

The proof is by induction on $p$:

- For $p = 0$, the claim holds trivially with $h_i = 0$ for all $i$.
- Suppose the claim holds for $p-1$. That is, we have for all $k \in \{1, \ldots, p-1\}$, $\overline{s^k}(g_1 + h_1, \ldots, g_n + h_n) \neq 0$, and $h_i \leq (p-1)^2$ for all $i$.

  If $\overline{s^p}(g_1 + h_1, \ldots, g_n + h_n) \neq 0$, we can choose the same numbers $h_1, \ldots, h_n$.

  If $\overline{s^p}(g_1+h_1, \ldots, g_n+h_n) = 0$, let $i_0$ be such that $i_0 \neq 0$ and $d^p_{i_0} \neq 0$, i.e. $d^p_{i_0}$ is a non-zero coefficient of a variable in $\overline{s^p}$. Such an index exists since $s^p$ is non-null and $\overline{s^p}(g_1 + h_1, \ldots, g_n + h_n) = 0$ implies that $s^p$ cannot be the constant term $d^p_0$. We can now replace the argument $g_{i_0} + h_{i_0}$ by $g_{i_0} + h_{i_0} + 1$, in which case $\overline{s^p}$ will evaluate to $d^p_{i_0}$, which is non-zero, as desired. The problem is that this replacement may nullify a function $\overline{s^{k_0}}$ with $k_0 < p$. Note that this is only possible if $d^{k_0}_{i_0} \neq 0$, i.e. $\overline{s^{k_0}}$ must have a non-zero coefficient at the same position $i_0$ as $\overline{s^p}$. To re-enforce that $\overline{s_{k_0}}$ evaluates to non-zero, we replace $g_{i_0} + h_{i_0} + 1$ by $g_{i_0} + h_{i_0} + 2$. This replacement does *not* nullify $\overline{s^p}$ again, as the following argument shows: For $k \in \{1, \ldots, p\}$, $i \in \{1, \ldots, n\}$, integers $a_1, \ldots, a_n$ and $h \in \mathbb{Z}$:

  $$\overline{s_k}(a_1, \ldots, a_{i-1}, a_i + h, a_{i+1}, \ldots, a_n) = \overline{s_k}(a_1, \ldots, a_n) + d^k_i \cdot h\,.$$

  In particular, if $\overline{s_k}(a_1, \ldots, a_n) = 0$ and $d^k_i \neq 0$, then replacing the argument $a_i$ by any *larger* integer results in a non-zero value of $\overline{s_k}$. Thus, to complete the inductive step, we increase $g_{i_0} + h_{i_0}$ until none of the functions $\overline{s_k}$ evaluates to 0. Since there are $p$ such functions, this requires at most $p$ increases (of magnitude 1), thus $h_{i_0} \overset{\text{(IH)}}{\leq} (p-1)^2 + p \leq p^2$. Note that values $h_i$ with $i \neq i_0$ are not affected; here we simply have $h_i \leq (p-1)^2 \leq p^2$.

This concludes the inductive proof. What remains to show is that, with $z_i := g_i + h_i$ for all $i$, we have $f(z_1, \ldots, z_n) \leq 0$:

$$\begin{aligned}
f(z_1,\ldots,z_n) &= \max_{j=1}^m \left(c_0^j + \sum_{i=1}^n c_i^j (g_i + h_i)\right) \\
&\overset{(*)}{\leq} \underbrace{\max_{j=1}^m \left(c_0^j + \sum_{i=1}^n c_i^j g_i\right)} + \sum_{j=1}^m \left|\sum_{i=1}^n c_i^j h_i\right| \\
&\leq \qquad f(g_1,\ldots,g_n) \qquad + \sum_{j=1}^m \sum_{i=1}^n |c_i^j| p^2 \\
&= \qquad f(g_1,\ldots,g_n) \qquad + p^2 \cdot ||C|| \\
&\overset{(**)}{\leq} (f(y_1,\ldots,y_n) + ||C||) \quad + p^2 \cdot ||C|| \\
&\overset{(***)}{\leq} \qquad (r + ||C||) \qquad + p^2 \cdot ||C|| \\
&= 0\,,
\end{aligned}$$

where $(*)$ applies property 16 ($m$ times), $(**)$ applies property 17, and $(***)$ is by the choice of $r$.

# E Proofs generating PAID+UF Interpolants

## E.1 Proof of Lemma 12: Sufficient Conditions for PAID+UF$_p$ Interpolants

The reasoning is similar as for Lem. 6 in Sect. C.1. Consider a sub-proof of $\mathcal{P}$ of the form shown in Lem. 12. We only consider the case $D = L, E = R, F = R$, as the other cases are similar. The interpolant generated by the sub-proof is

$$J_6 \;=\; \exists_{L\bar{s}_1 t_1 \bar{s}_2 t_2}\, J_5 \;=\; \exists_{L\bar{s}_1 t_1}\, J_5 \;=\; \exists_{L\bar{s}_1 t_1} \big(J_3 \wedge f_p(\bar{s}_1, t_1) \wedge J_4\big)$$

By definition of PAID+UF$_p$, we know that $t_1$ is a Skolem constant. If $t_1$ is $L$-local at this point in the proof (the quantifier $\exists_{Lt_1}$ does not disappear), then it is $L$-local also in $\mathcal{Q}$, which means that $t_1$ does not occur in the interpolant $J_3$. This implies:

$$J_6 \;=\; \cdots \;\equiv\; \exists_{L\bar{s}_1} \big(J_3 \wedge\, \exists_{Lt_1}\, (f_p(\bar{s}_1, t_1) \wedge J_4)\big)$$

Furthermore, observe that the constants quantified by $\exists_{L\bar{s}_1}$ are $L$-local in $\mathcal{R}$, so that none of them occurs in $J_4$. As in the proof of Lem. 6, the expression $\exists_{L\bar{s}_1} \big(J_3 \wedge \cdots\big)$ can then be transformed to a sequence of guarded quan-

tifiers:

$$
\begin{aligned}
J_6 \;\equiv\; & \exists y_1.\ \big(\alpha\beta_1 y_1 + (L^{-1}\bar{u}')_1 \doteq 0 \wedge \\
& \exists y_2.\ \big(\alpha\beta_2 y_2 + (L^{-1}\bar{u}')_2 \doteq 0 \wedge \\
& \quad\vdots \\
& \exists y_r.\ \big(\alpha\beta_r y_r + (L^{-1}\bar{u}')_r \doteq 0 \ \wedge\ \exists_{Lt_1}\ (f_p(LS\bar{y} + \bar{v}, t_1) \wedge J_4)\big) \cdots \big) \\
& \wedge \bigwedge_{i=r+1}^{k} (L^{-1}\bar{u}')_i \doteq 0
\end{aligned}
$$

To conclude the proof, we have to consider two cases:

– $t_1$ is $L$-local and the quantifier $\exists_{Lt_1}$ does not disappear: then we are finished, because it has been shown that $J_6$ is equivalent to a PAID+UF$_p$ formula.
– $\exists_{Lt_1}$ disappears: in this case, we can rewrite the formula $f_p(LS\bar{y} + \bar{v}, t_1) \wedge J_4$ to $\exists z.\ (f_p(LS\bar{y} + \bar{v}, z) \wedge t_1 \doteq z \wedge J_4)$, which is in PAID+UF$_p$.

### E.2 Proof of Theorem 10: Closure of PAID+UF under Interpolation

Most importantly, we can first observe that a completeness result similar to Thm. 5 also holds for PAID+UF$_p$ (given in the next lemma). Theorem 10 then follows as a simple implication, because PAID+UF formulae can be translated to PAID+UF$_p$, interpolated, and the interpolant translated back to PAID+UF.

**Lemma 18 (Completeness).** *Suppose that $A_{Rel}, B_{Rel}, \{FC_f\}_{f \in F_A \cup F_B} \vdash \emptyset$ is valid. Then there is a formula $I$ such that (i) the sequent*

$$\lfloor A_{Rel} \rfloor_L, \lfloor B_{Rel} \rfloor_R, \{\lfloor FC_f \rfloor_L\}_{f \in F_A}, \{\lfloor FC_f \rfloor_R\}_{f \in F_B} \vdash \emptyset \ \blacktriangleright I \qquad (16)$$

*is provable in the calculus of Sect. 3.1, enriched with the rules* ALL-LEFT-GRD *and* EX-RIGHT-GRD*, and (ii) $I$ is a PAID+UF$_p$ formula up to normalisation of guards in expressions* (2)*.*

*Proof.* Given a sequent such that $A_{Rel}, B_{Rel}, \{FC_f\}_{f \in F_A \cup F_B} \vdash$ is valid, we can construct a proof of (16) satisfying the conditions in Lem. 12 using a procedure similar to the one in Sect. C.2. Lem. 12 then guarantees that the interpolant $I$ is in PAID+UF$_p$ up to guard normalisation.

In the procedure of Sect. C.2, only step 3 has to be changed to obtain an algorithm for PAID+UF$_p$: instead of searching for complementary literals $p_k(\bar{u}_k)$ and $p_m(\bar{u}_m)$, in the PAID+UF$_p$ case we have to check for literals $f_p(\bar{s}_1, t_1)$ and $f_p(\bar{s}_2, t_2)$ such that $\bar{s}_1 \doteq \bar{s}_2$ is implied by the equalities in the antecedent. If such a pair has been detected, the consistency axiom $FC_f$ can be instantiated, closing the first three premises as dictated by Lem. 12. For the fourth premise, the proof procedure can go back to step 2 and continue proving. To ensure termination of the overall procedure, it only has to be guaranteed that the axiom $FC_f$ is not repeatedly instantiated on the same branch for the same pair of literals $f_p(\bar{s}_1, t_1)$ and $f_p(\bar{s}_2, t_2)$. □